\documentclass[acmtog]{acmart}

\def\BibTeX{{\rm B\kern-.05em{\sc i\kern-.025em b}\kern-.08emT\kern-.1667em\lower.7ex\hbox{E}\kern-.125emX}}
    
\begin{document}

%
\title{Online Motion Style Transfer for Interactive Character Control}

%
\author{Yingtian Tang}
\email{yingtian@seas.upenn.edu}
\affiliation{%
  \institution{University of Pennsylvania}
}

\author{Jiangtao Liu}
\email{liujt14@gmail.com}
\affiliation{%
  \institution{The Chinese University of Hong Kong}
}

\author{Cheng Zhou}
\email{mikechzhou@tencent.com}
\affiliation{%
  \institution{Tencent Robotics X}
}

\author{Tingguang Li}
\email{tgli0809@gmail.com}
\affiliation{%
  \institution{Tencent Robotics X}
}

%

%
\begin{abstract}

Motion style transfer is highly desired for motion generation systems for gaming. Compared to its offline counterpart, the research on online motion style transfer under interactive control is limited. In this work, we propose an end-to-end neural network that can generate motions with different styles and transfer motion styles in real-time under user control. Our approach eliminates the use of handcrafted phase features, and could be easily trained and directly deployed in game systems. In the experiment part, we evaluate our approach from three aspects that are essential for industrial game design: accuracy, flexibility, and variety, and our model performs a satisfying result.

\end{abstract}

%
%

\ccsdesc[500]{Computing methodologies~Motion processing}
\ccsdesc[300]{Computing methodologies~Neural networks}

%
\keywords{motion synthesis, style transfer, interactive control}

%

%
\maketitle

\section{Introduction} \label{sec:intro}

Character motion synthesis is a major part of game design. The generation of character motions requires natural joint placements and fluent movements. In video games, there are multiple styles of character motions that correspond to various personalities, ages, and \textit{etc.} For example, there might be a monster who walks in a zombie-like manner, or a warrior who runs while holding a weapon. 

Traditional methods for motion synthesis \cite{kovar2008motion,buttner2015motion} construct a database of motion clips. At runtime, the algorithm searches in the database for the best match, according to the user control and the current character motion, and then interpolates to generate the motion for the next frame.
These frameworks consider different styles separately and record different motion clips for them. 
As a result, they require large amounts of motion capture data, and the generated motions are restricted to the preset styles.

Data-driven methods empowered the recent developments for character motion synthesis, of which the neural network methods played a critical role. A series of studies \cite{holden2017phase,zhang2018mode,starke2019neural,starke2020local} initiated the exploration of the application of neural networks in motion synthesis. These methods avoid searching through a large database, but still producing character motions with high quality in various scenarios, such as quadruped motions \cite{zhang2018mode} and character interactions \cite{starke2020local}. These motion generation models mostly possess a specialized architecture called Mix-Of-Expert (MOE) to modulate their model parameters according to the phase (timing of the cyclic movements) and other properties of the current motion. These model structures are suitable for online motion generation with user control. However, in this case, a single model cannot produce motions with different styles, so that multiple models have to be trained. 

There are multiple research studies \cite{aberman2020unpaired,smith2019efficient,holden2017fast,yumer2016spectral} on the topic of motion style transfer. In motion style transfer, the original style of a given motion clip could be transferred to another style, where the content of that motion clip (for example, walking) is preserved. Nevertheless, these methods only focused on motion style transfer in an offline fashion, which means they transferred the style of a static motion clip. In this work, we study the methods for online motion style transfer with interactive user controls, which could be directly deployed in game systems. With per-frame user controls, the algorithm generates new stylized motions on the fly, in an autoregressive fashion. A similar work on this topic is \cite{mason2022real}.

For game design, the styles provide diversity for the designers. Motions with different styles could be used for different characters, or splitting and stitching to create impressive visual effects. Based on this, we design an evaluation pipeline and focus on three aspects of the model: the \textit{accuracy}, the \textit{flexibility}, and the \textit{variety}. 

First, the model should \textit{accurately} reproduce the original motions in the training data, given the same controls. This ensures that following different paths, the model generates diverse motions learned from the datasets. Second, the model should \textit{flexibly} make online transition between styles according to the control, which further enhances the diversity. For example, this empowers a character to have actions with different respective style. Third, the model should \textit{create new styles} by merging existing styles. This makes the stylized motions not limited to preset ones. With these considerations, we device both qualitative and quantitative evaluations, as will be shown in Sec.~\ref{sec:exp}. 

We propose the Motion Temporal Convolution Network (MTCN) for multi-style online motion synthesis with interactive controls.
A single MTCN model can be directly used in games for generating a character with multiple styles and actions, according to the user control. Moreover, the model performs well on all three aforementioned style-related functionalities.

Our proposed model has two essential augmentations on the previous online motion generation models \cite{holden2017phase,zhang2018mode,starke2019neural,starke2020local}. The first is to re-modulate the MOE with the style. The MOE architecture is modulated by a set of parameters called expert weights, which is further re-modulated in our model by the assigned style. The expert weights change according to different style assignments. The second is to replace the phase with learned features. The phase of motions is a key input to generate expert weights in the previous models. In the recent work \cite{starke2020local}, the phase is computed through fitting a sinuous curve to the foot/hand contact labels. We replace the phase input with Temporal Convolution Network (TCN) motion features, which better accommodate complex and multi-style motions.

In summary, we have the following contributions:

\begin{itemize}
    \item We propose an end-to-end model for online and interactive motion style transfer. With our model, the user could control the game character to switch between styles or to switch to a merged style fluently and naturally. The model could be directly deployed in game systems.
    \item We replace the commonly used handcrafted phase with learned features, which make the model better suited to multi-style tasks and has simpler data-preprocessing procedure. 
\end{itemize}

The later sections are structured as follows: Section ~\ref{sec:related work} reviews the development of MOE-based models and the advance of offline motion style transfer. Then, Section ~\ref{sec:method} demonstrates how we modify the MOE so that it is conditioned on the style assignment, and then proposes our method MTCN which is based on learned TCN motion features. The data preparation is described in Section ~\ref{sec:data}. In Section ~\ref{sec:exp}, we compare the MTCN with the baseline model, in terms of the accuracy, the flexibility, and the variety. Finally, Section ~\ref{sec:conclusion} concludes the paper and indicates our future work.

\section{Related Works} \label{sec:related work}

In this section, we review the series of neural network methods that utilized the MOE architecture for online motion synthesis, as well as the style transfer approaches in image and motion generation. Later in Sec.~\ref{sec:method}, we will demonstrate how the style transfer techniques could be combined with the MOE for online multi-style motion synthesis.

\subsection{Mix-Of-Expert for Online Motion Synthesis} \label{sec:moe}

The series of neural network methods \cite{holden2017phase,zhang2018mode,starke2019neural,starke2020local} use the Mix-Of-Expert (MOE) architecture for accommodating different phases, controls, and interactions in motion synthesis. Denote the character motion (i.e., the joints positions and rotations in the local coordinate) at time $t$ as $m_t$, these works formulate the synthesis task as a sequence generation task. That is, to learn a mapping $m_{t+1} = \psi (g_t, m_t, m_{t-1}, ...)$, where $g_t$ is the user control given at time $t$. More specifically, they claimed that an autoregressive formulation $m_{t+1} = \psi (g_t, m_t)$ would benefit the performance.

A simple realization of the mapping $m_{t+1} = \psi (g_t, m_t)$ is to directly use a feed-forward neural network and train it with regression losses. However, \cite{holden2017phase} argued that this led to sliding movements. They had the investigation that the network parameters should be modulated according to the phase of the motions (denoted as $p_t$), which refers to the cyclic left-right feet movements in their case. Specifically, it is a periodic variable ranging from 0 to 1, determined by the cyclic movement of feet. Then they proposed the MOE structure and used it in their Phase-Functioned Neural Network (PFNN). The MOE is conditioned on a set of "expert weights", denoted by $\boldsymbol{\alpha}=\{\alpha_n | n=1,2,...,N_A, \sum_{n} \alpha_n = 1, \alpha_n \geq 0 \}$, where $N_A$ is the total number of experts. For a neuron in layer $i$, the forward-propagation is changed from the original $y=\phi(w_i*x+\beta_i)$ to: 
\begin{equation} \label{eq:moe_layer}
     y= \phi \left [ ~ \sum_{n=1}^{N_A}(w_{n} * \alpha_n)x + (\beta_n * \alpha_n) ~ \right ]
\end{equation}

\noindent where $w_n$ and $\beta_n$ are the network weight and bias for different experts. $\phi$ is the non-linear activation function. Compared with the simple linear layer, the structure of MOE parameters could naturally possess stronger capacity. With the change of conditional parameters $\alpha_n$, the MOE produces continuously changing parameters for networks. The constraints $\sum_{n} \alpha_n = 1$ and $\alpha_n \geq 0$ are required, out of the intuition that the combination of experts should be convex.

The network weights of PFNN cyclically change according to the phase  value $p_t$, which is generated by a phase function. The subsequent studies further developed this idea. \cite{zhang2018mode} proposed to use the velocity of the current motion to replace the the simple phase $p_t$, so that the MOE can adapt to different modes of quadruped motions. \cite{starke2019neural} developed the Neural State Machine (NSM) model that takes $p_t$ along with action labels and environment interaction variables to produce $\boldsymbol{\alpha}$. \cite{starke2020local} proposed to use multiple local phases instead of a single global phase to capture complex interactions involving multiple contacts. 


We choose NSM as our baseline model, since it has strong capacity and does not consider multiple contacts, which is our case. In this work, we mainly consider locomotion data, i.e. standing, walking, and running, so that we discard the environment interaction variables in the original NSM. 

The inputs of the NSM model at time $t$ consist of four components, namely the joint positions and rotations $m_t$, root trajectory $r_t$, gait trajectory $\lambda_t$ and the movement phase $p_t$, such $\boldsymbol{X_t} = \{ m_t, r_t, \lambda_t, p_t \}$. 

Specifically, to produce $\boldsymbol{\alpha}$ for MOE, NSM used a multi-layer network (called gating network) that takes as input the Kronecker Product of the phase $p_t$ and the gait trajectory $\gamma_t$ ($p_t \otimes \gamma_t$). We denote the root trajectory at $t$ as $r_t$. The two trajectories, $\gamma_t$ and $r_t$, consist of the sampled gaits (for example, stand, walk, or run) and root positions (the positions of past and future local coordinates, related to the current local coordinate) of the character from the past 1s to the future 1s. The sampling rate is $1/6$ seconds. We can modify $r_t$ to control to which direction the character moves. 

Controlled by expert weights $\boldsymbol{\alpha}$, the MOE then takes the ${{m}_t}, {r}_t, \gamma_t$ to regress the ${{m}_{t+1}}, {r}_{t+1}, \gamma_{t+1}$. Denote the gating network as $\mathcal{G}$, and the MOE as $\mathcal{M}$, the overall NSM model is shown in Eq.~\ref{eq:nsm}. 
\begin{equation} \label{eq:nsm}
\begin{split}
     \boldsymbol{\alpha} & = \mathcal{G} (~ p_t \otimes \gamma_t~ ) \\
    {m}_{t+1}, {r}_{t+1}, \gamma_{t+1} & = \mathcal{M} ( {m}_{t}, {r}_{t}, \gamma_{t} ; \boldsymbol{\alpha}  )   
\end{split}
\end{equation}

Here, for discussing style-related operations, we shall expand the description of a motion ${m}$ to ${m}^s$, with a style $s \in \mathcal{S}$. We shall also denote a sequence of motion as ${m_{t_0,t_1}^s}$ which contains the motions from $t_0$ to $t_1$.

These models based on the MOE realize ${m^s_{t+1}} = \psi (g_t, {m^s_{t}})$, and generate motions with high quality for different scenarios and the same style. In this work, we extend this mechanism for multi-style online motion generation, which is ${m^{s'}_{t+1}} = \psi (g_t, {m^s_{t}}, s')$, with an online target style $s' \in \mathcal{S}$.

\subsection{Motion Style Transfer} \label{sec:motion style transfer}

Unlike the online motion style transfer tasks, the offline motion style transfer is provided with the whole extracted content sequence $c_{t_0,t_1}$, and generates ${m^{s'}_{t_0,t_1}}$ with the target style $s'$ given.

This kind of offline style transfer has been well-studied in the field of computer vision. \cite{gatys2016image,johnson2016perceptual} started the research of image style transfer using neural networks, by noticing that the style of an image could be described by the inter-layer feature statistics of a pre-trained Convolutional Neural Network (CNN). \cite{ulyanov2016instance} further showed that the second order statistics of the aforementioned inter-layer feature distributions are enough for style descriptions. They proposed an Instance Normalization (IN) which normalizes the means and variances of inter-layer feature distributions, so that the style of the image is discarded and the content is preserved. With IN, their image style transfer outperformed the previous methods. \cite{huang2017arbitrary} realized that a de-normalization operation right after IN with a pair of learned mean and variance (or standard deviation) could finish style transfer process. Their method named adaptive Instance Normalization (AdaIN), learned different pairs of mean and variance for different styles, which achieved arbitrary image style transfer.

To briefly summarize, the change or elimination of styles of images is directly related to the manipulation of inter-layer feature distributions in the neural networks.

Recently, \cite{aberman2020unpaired} applied the concept of content and style of images on motion clips. The original 2-D convolution kernels in a normal CNN is replaced by 1-D convolution kernels, which change the CNN into a TCN. They extracted the content sequence from a motion clip using an TCN encoder $\boldsymbol{E_c}$ with Instance Normalization (IN, \cite{ulyanov2016instance}), and also extracted the style with a TCN encoder $\boldsymbol{E_s}$. Then, they applied AdaIN on a TCN decoder $\boldsymbol{G}$ to generate unseen content-style combination, with Generative Adversarial Loss \cite{goodfellow2014generative}. Overall, their method is formulated as the following:
\begin{equation} \label{eq:sgan}
\begin{split}
     c_{t_0,t_1} = &~ \boldsymbol{E_c}(m^s_{t_0,t_1}) \\
     s' = &~ \boldsymbol{E_s}(m^{s'}_{t_0,t_1}) \\
     {m^{s'}_{t_0,t_1}} = &~ \boldsymbol{G}(c_{t_0,t_1}, s') 
\end{split}
\end{equation}

However, their work only focused on this kind of offline whole-sequence motion style transfer, while for real game systems the online counterpart would be much more desired.

\section{Method} \label{sec:method}

\begin{figure}[t!]
\centering
\includegraphics[width=\linewidth]{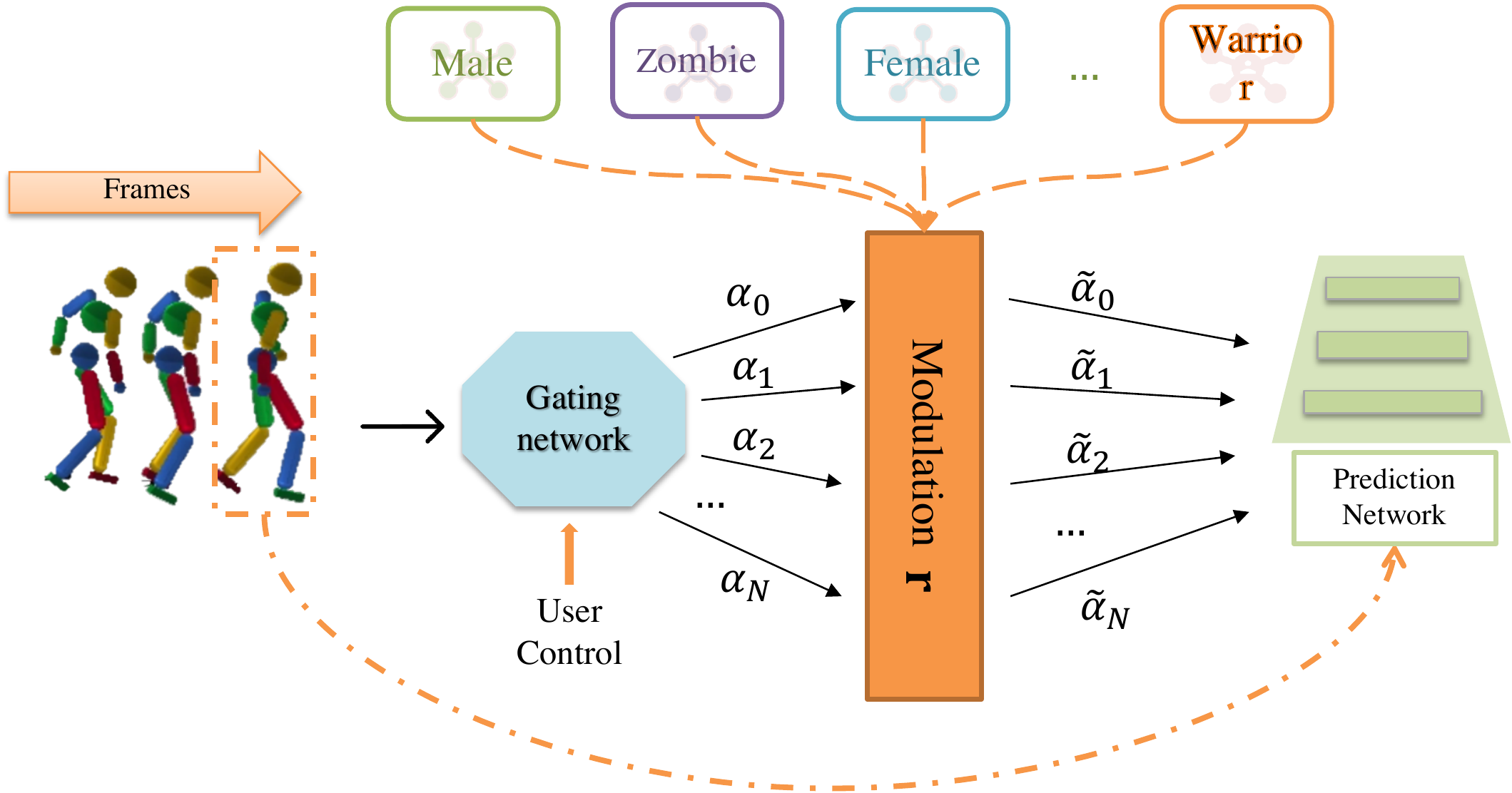}
\caption{\small{The Style-NSM is a modified version of NSM. It takes the generated expert weights $\boldsymbol{\alpha}$ and modulate them according to the style $s$(as shown in the dashed box part in the above image). Then, the modulated expert weights are used to control the MOE.}}
\label{fig:snsm}
\end{figure}

In this section, we propose our method Motion Temporal Convolution Network (MTCN) for multi-style online motion synthesis. Before introducing MTCN, we first demonstrate how the MOE mechanism (introduced in Sec.~\ref{sec:moe}) could be combined with the ideas from motion style transfer (introduced in Sec.~\ref{sec:motion style transfer}). This direct combination leads to Style-NSM (SNSM), which will be our baseline for developing MTCN. MTCN further uses the learned TCN motion features to replace the phase input $p_t$ in NSM, which results in better performance on multi-style online motion generation.

\subsection{Style-NSM} \label{sec:snsm}

As mentioned before, the MOE is an important mechanism for generating motions in an online manner. The expert weights change according to the phase and gait trajectory, adjusting the whole network weights, as is reflected in Eq.~\ref{eq:moe_layer}. This equation could be also re-written as the following:
\begin{equation} \label{eq:moe_layer_rewrite}
     y= \phi \left [ ~ \sum_{n=1}^{N_A}(w_{n}x + \beta_n) * \alpha_n ~ \right ]
\end{equation}

In this form, the MOE could be regarded as modulating the outputs of several neurons and then summing them up. Here the modulation operation is exactly the same as it is in the de-normalization operation in AdaIN. More specifically, it changes the variance of the inter-layer feature distribution. With this observation, we can see the similarity between MOE and AdaIN.

We can further take the advantage of this property, and re-modulate the expert weights by style label so that they also adapts to different styles. 
We change the forward-propagation for each layer from Eq.\ref{eq:moe_layer} to the following:
\begin{equation} \label{eq:moe_relax}
     y= \phi \left [ ~  \sum_{n=1}^{N_A}(w_{n}x + \beta_n) * r ( \alpha_n , s ) ~ \right ] 
\end{equation}
\noindent where $s$ is the style specification and $r$ is the modulation function. There are multiple choices of $r$, here we use a simple element-wise affine transformation, which is the following:
\begin{equation}
     r ( \alpha_n , s ) = \sigma (s) * \alpha_n + \mu(s)
\end{equation}
\noindent where $\sigma (s)$ and $\mu (s)$ are neural nets which take the one-hot encoding of style $s$. The reason of this choice is that this architecture is empirically more stable than the neural net counterpart, and it introduces minimal number of extra parameters. For the modulation to work better, we shall remove the constraint imposed on expert weights, that is, $\sum_{n} \alpha_n = 1$ and $ \alpha_n \geq 0 $. Then, the $\alpha_n$ should take arbitrary real values and so do the modulated results. 

Overall, the SNSM model is different from NSM with an additional modulation to generate the expert weights (the comparison is shown in Fig.~\ref{fig:snsm}):
\begin{equation} \label{eq:snsm}
\begin{split}
     \boldsymbol{\alpha} & = r ( \mathcal{G} (~ p_t \otimes \gamma_t~ ) , s' )
\end{split}
\end{equation}

\begin{figure}[t!]
\centering
\includegraphics[width=\linewidth]{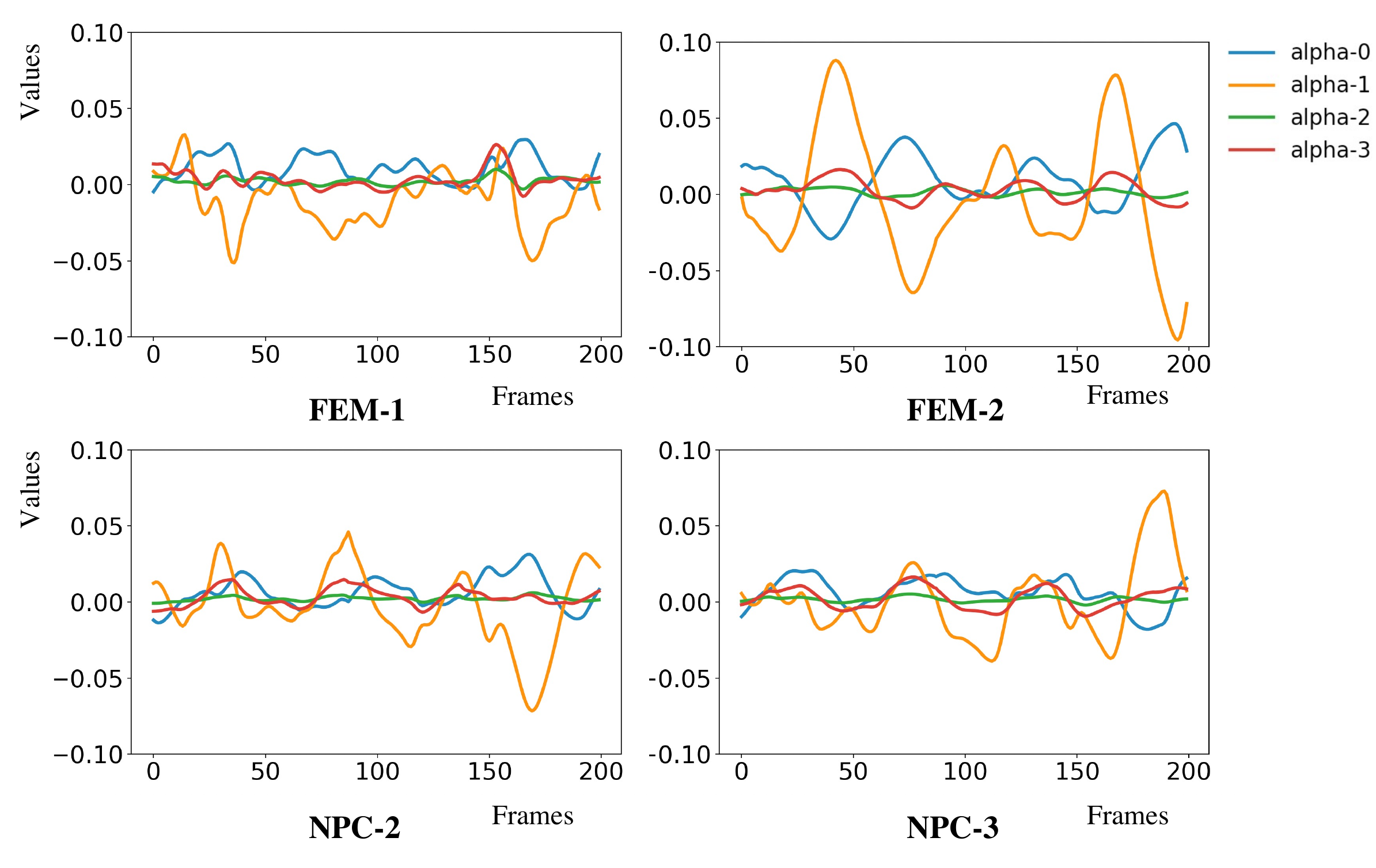}
\caption{\small{In SNSM, the expert weight blending coefficients (the four ``alpha''s values) of four different styles are shown here. Each plot visualizes four weights' values of one specific style, where the horizontal axis indicates the frame range.}}
\label{fig:experts}
\end{figure}

The SNSM thereby implements the mapping $\boldsymbol{m^{s'}_{t+1}} = \psi (g_t, \boldsymbol{m^{s}_{t}}, s')$. There is a concern that how the original style $s$ is removed so that the new style $s'$ could take place. Empirically, we find that the outputs of autoregressive models have small dependency on the original style of $\boldsymbol{m^{s}_{t}}$. When there is an online style transition, we interpolate between the original and the target style to gradually make the transition. This process is also shown in Fig.~\ref{fig:experts}.

\subsection{Motion Temporal Convolution Network} \label{sec:mtcn}

\begin{figure*}[t!]
\centering
\includegraphics[width=.9\linewidth]{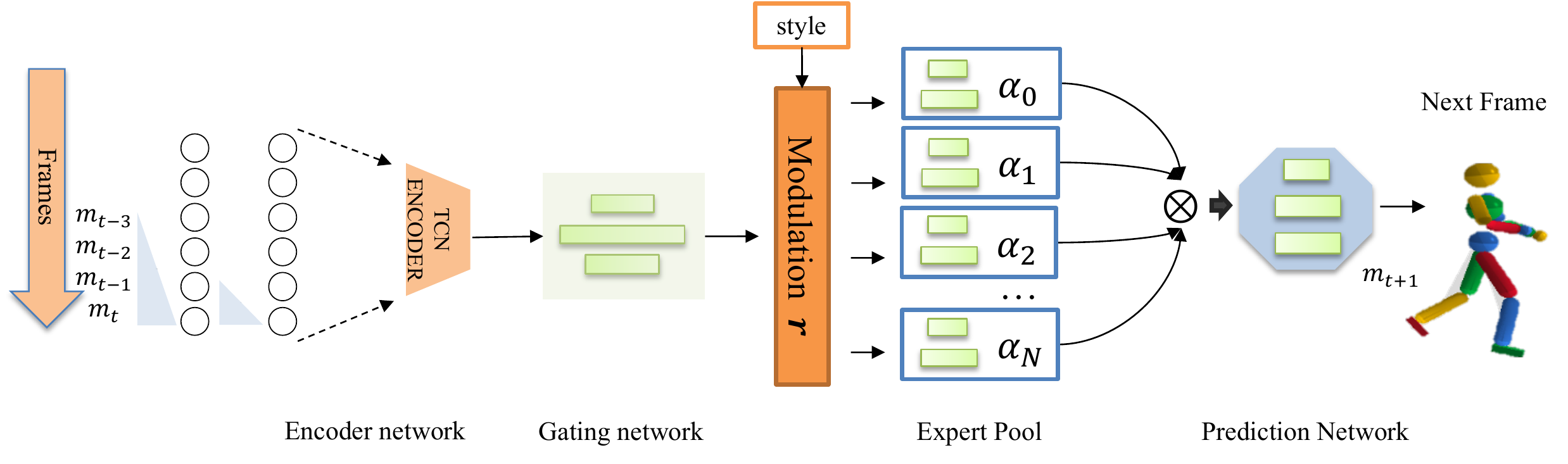}
\caption{\small{The MTCN takes the past sequence of motions to generate expert weights, but still uses the MOE in an autoregressive fashion. MOE only takes the last frame of motion as the input.}}
\label{fig:mtcn}
\end{figure*}

The NSM inputs the phase $p_t$ and gait $\gamma_t$ to the gating network to generate the expert weights. The idea of using $p_t$ was first described in PFNN \cite{holden2017phase}. Indeed, we empirically find that $p_t$ is a critical input variables for enabling the MOE for online motion synthesis. However, a simple handcrafted global phase has limited expressive capability. For some stylized motions, for example, the zombie-like walking, the phase of the motions is complex and irregular. Besides, the handcrafted process adds to the complexity of the whole system. As a result, we propose to use a Temporal Convolutional Network (TCN) for direct end-to-end feature extraction. The extracted motion features can replace the original handcrafted phase inputs.

Since we target at online motion synthesis, we have to make the TCN causal, which means it does not take the future information as inputs. The common form of TCN is to directly apply symmetric kernels along the time dimension, \textit{i.e.,} $v^{l+1}_{t} =\sum_{k=-\tau}^{\tau} v^{l}_{t-k} w_{k}$. ($v_t^l$ is the value of channel $l$ at time $t$, $w$ is the kernel weight, and $\tau$ is the scope length) In our case, it is more natural to limit the scope of the kernel so that it only takes the past inputs. So, we instead use the Causal Convolution proposed in \cite{oord2016wavenet}, which takes the form of $v^{l+1}_{t} =\sum_{k=0}^{2\tau} v^{l}_{t-k} w_{k}$. 

Replace the inputs to the gating network of SNSM with the proposed motion features extracted by a TCN encoder $\mathcal{T}$, we have the expert weights of our new model MTCN to be:
\begin{equation} \label{eq:mtcn}
\begin{split}
     \theta_t &= \mathcal{T} ({{m}^s_{t-\tau,t}}) \\
     \boldsymbol{\alpha} & = r ( \mathcal{G} ( \theta_t ) , s' )
\end{split}
\end{equation}

\noindent where $\theta_t$ is the motion features extracted from the motion. ($\tau$ is a preset window size) To ensure the modulation capacity, the gating network generates $\alpha_n^l$ for each layer $l$ separately, \textit{i.e.,} $\boldsymbol{\alpha}=\{ \alpha_n^l, n=1,2,...,N_A, l=1,2,...,L \}$, instead of generating $\alpha_n$ that are shared by all layers, where $L$ is the total number of layers of MOE. As shown in Fig.~\ref{fig:mtcn}, at each frame, the MTCN takes the past sequence to generate TCN motion features, which determine the expert weights. Then, the expert weights are re-modulated according to the assigned style, which finally controls the MOE.

\textit{Scheduled Sampling.} We find that Scheduled Sampling \cite{bengio2015scheduled} is a crucial tool for training MTCN. Scheduled Sampling aims to expose the regression error at runtime to the model at training time. More specifically, for each training iteration, a sampling probability $p$ is chosen to decide whether the inputs of the next frame are the real data or they are the outputs from the network for the previous frame. With probability $1-p$, the inputs become the outputs from the network, which can be regarded as samples from runtime distribution. This sampling process takes place for multiple frames, and the errors are accumulated and backpropagated at once. In practice, we gradually increase $p$ from 0 to 1 within the first 10 training iterations.

\textit{Window Instance Normalization.} With the introduction of the temporal motion features, the network has much more temporal dependency on past styles compared to the fully autoregressive SNSM. This is also confirmed in our experiments as we will show in Sec.~\ref{sec:exp}. Consequently, we also find it necessary to add the Instance Normalization (IN) to the encoder $\mathcal{T}$, so it becomes $\mathcal{T}_{IN}$. We refer to the MTCN after this modification as MTCN-IN. However, there is a subtle problem regarding the behavior of IN at runtime. As mentioned, IN works by taking a set of instances, in our case, a sequence of motions. At runtime, the motions are generated continuously. At time stamp $T$, it is undesirable to feed all the motions from the beginning to $T$ to the IN, because this whole sequence might already contain multiple styles and different motion types, which clearly differs from the short and single-styled motion sequences when training. Here we proposed a Window Instance Normalization (WIN) to handle this problem. Basically, the WIN takes sequences of motions within a pre-defined window size, $\tau$. At runtime, instead of normalizing over motions from beginning to $T$, WIN normalizes motions from $T-\tau$ to $T$. For a given sequence of motion ${m}_{{t_0,t_1}}^{s}$, for a specific intermediate layer, the neural net produces feature sequence ${f}_{{t_0,t_1}}^{s}$. The WIN takes place by:
\begin{equation} \label{eq:win}
\begin{split}
     \mu &= \frac{\sum_{t=t_0}^{t_1} {f}_{t}^{s}}{t_1 - t_0 + 1} \\
     \sigma &= \sqrt{ \frac{\sum_{t=t_0}^{t_1} ({f}_{t}^{s}-\mu)^2}{t_1 - t_0}} \\
     \widetilde{f}_{t}^s & = \frac{{f}_{t}^s - \mu}{\sigma} , ~for~all~t \\
\end{split}
\end{equation}

\noindent where the $s$ information is reduced in $\widetilde{f}_{t_0,t_1}^s$. After several WIN operations, the style $s$ of ${{m}^{s}_{t_0,t_1}}$ is largely reduced and finally produces the content sequence $c_{t_0,t_1}$.  

\textit{Modification of IN \& WIN.} There is a critical issue that applies to both IN and WIN. Consider the standing action, the character stands still on the ground with some very small yet natural movements. For Instance Normalization, these natural tiny movements are also normalized, thus generating undesirable content signals. Ideally, these movements should be represented with their original small values, instead of the normalized values. In fact, the standing pose is not only the case where this phenomenon happens. The consequence of these is the generated motions are shaking at runtime, instead of standing still or keeping the pose. As a result, we change the standard normalization procedure (last line in Eq.\ref{eq:win}) by imposing a minimum for standard deviation:
\begin{equation} \label{eq:win_mstd}
\begin{split}
     \widetilde{f}_{t}^s & = \frac{{f}_{t}^s - \mu}{max(\sigma, \epsilon)} , ~for~all~t \\
\end{split}
\end{equation}

\noindent where $\epsilon$ is the minimal value of standard deviation $\sigma$. Movements that are essentially small remain their small-value property after the WIN, such as standing, while other movements are normalized as usual. In practice, the choice of $\epsilon$ is set to be $0.3$, but generally works well within an empirical range of $0.1 \sim 0.3$.

\section{DATA PREPARATION} \label{sec:data}

The data preparation and character skeleton are described in this section, including the motion capture, the data labelling and bone re-targeting. 

\subsection{Motion Capture and Labelling}  \label{sec:capture and label}
We collect data with different styles. The whole data consist of 8 styles in total, including three different zombie styles (NPC-1, NPC-2, and NPC-3), a gun-holding style (NPC-Gun), and four distinct female styles (FEM-1, FEM-2, FEM-3, FEM-4). The three zombie styles include a common zombie, a collapsing zombie, and a jumpy zombie. The four female styles cover styles of being injured, feeling cold, normal, and sexy. For each style, there are two separate BVH files that respectively record clockwise and counterclockwise circular walks in that style. In total, there are 16 BVH files and equivalently 40383 frames for the training. For all data, the bones are calibrated to be the same, and the exact same preprocessing pipeline is used, which include extracting root and gait trajectories as well as the phase computation. 

Each frame of data is annotated with a global phase value calculated by manually labelled footstep of four joints, "right angle", "right toe", "left angle" and "left toe". The periodically varied phase value describes the motion condition at current frame. Apart from the phase labels, we also label a action type for each frame,"stand" and "walk", which is the motion type of the character.

\subsection{Bone re-targeting}  \label{sec:re-target}
The character we used is similar to the Humanoid used by \cite{peng2018character}, it has 18 joints including the root joint set as base coordinate. To fit to a general representing and animation platform, we need to re-target the captured motion sequences, which have 48 joints.
The scheme is composed of the following three steps:
\begin{itemize}
    \item The joints size is firstly scaled to target joints of character, according to labelled skeleton. In details, we fit a template bones into positions and heights of the original motion sequences frame by frame. 
    \item The scaled "body" joints were re-targeted using dampened least squares inverse jacobian method. We set the damping parameter to be 2.0 to get fairly stable and effective end effectors after 20 iterations.
    \item The last step is to adjust the toe height to make the character keep contact with the ground with manually calculated value. 
\end{itemize}

After the labeling and re-targeting process, for the final prepared training set we find it has stable and smooth movement. The above mentioned process method can adapt to a wide range of characters without concerning the difference of motion capture systems. 

\section{Evaluations} \label{sec:exp}

As mentioned in Sec.~\ref{sec:intro}, we demand the model to have three properties, \textit{i.e.,} the accuracy, the flexibility, and the variety, for providing the diversity for game design. We therefore propose three corresponding evaluations for the online motion style transfer task: 
\begin{enumerate}
    \item Stylized Motion Replay: given the original control sequences, the model should accurately regenerate the training data for all styles.
    \item Online Style Transfer: the model should transfer to the assigned style by the user at runtime.
    \item Style Interpolation: the model should merge different learned styles at runtime.
\end{enumerate}

In this section, we show the results of all of the three evaluations. We use SNSM as our baseline, as it is the most straightforward modification of the NSM that enables multi-style motion generation. We then demonstrate the superiority of our proposed model MTCN. All models are set to have 4 expert weights and layers with 256 hidden units. The models are trained with batch-size 64 and Adam optimizer \cite{kingma2014adam} (with weight decay $1 \times 10^{-5}$ and initial learning rate $1 \times 10^{-3}$) for 100 iterations. When training, the value of dropout \cite{JMLR:v15:srivastava14a} is set to be 40\%.

It should be noticed that the first evaluation is the most important, since the accuracy of motion reproduction is strongly associated with the online motion generation performance and is the foundation for other style-related operations.

\subsection{Stylized Motion Replay}  \label{sec:replay exp}

Currently there are no widely used quantitative measures of the accuracy of a learned motion generation model. The stylized motion replay measure we used here is empirically a reasonably good indicator of the stylized motion quality in our experiments. 

While in the training we use Scheduled Sampling with a fixed window of prediction, we require the model to reproduce the exact same motion sequences at evaluation time, given the original control sequences. 

More specifically, at evaluation time, we iteratively input the pre-computed control signals $\gamma_t$ and $r_t$ to the model ($t$ starts from the beginning of the Biovision Hierarchy (BVH) file), and thus run the autoregression until the end. Then, we compute the Squared Error (SE) at each frame between the prediction ${{m}_{t}}^{s}$ and the ground truth (denoted as $\boldsymbol{\hat{m}^{s}_{t}}$). We then have a sequence of SEs, which is $e(t) = \|{{m}_{t}}^{s}-\boldsymbol{\hat{m}^{s}_{t}}\|^2$. To ensure that various motions of all the styles are accurately learned, we evaluate the models on all its training data, which include different styles and contents. We also average the errors by $(\sum_{t=1}^{T} e(t)) / T$ ($T$ is the total number of frames) for a BVH to get a Mean Squared Error (MSE), which is an overall error measure for the performance on that BVH.

\begin{figure}[t!]
\centering
\includegraphics[width=1\linewidth]{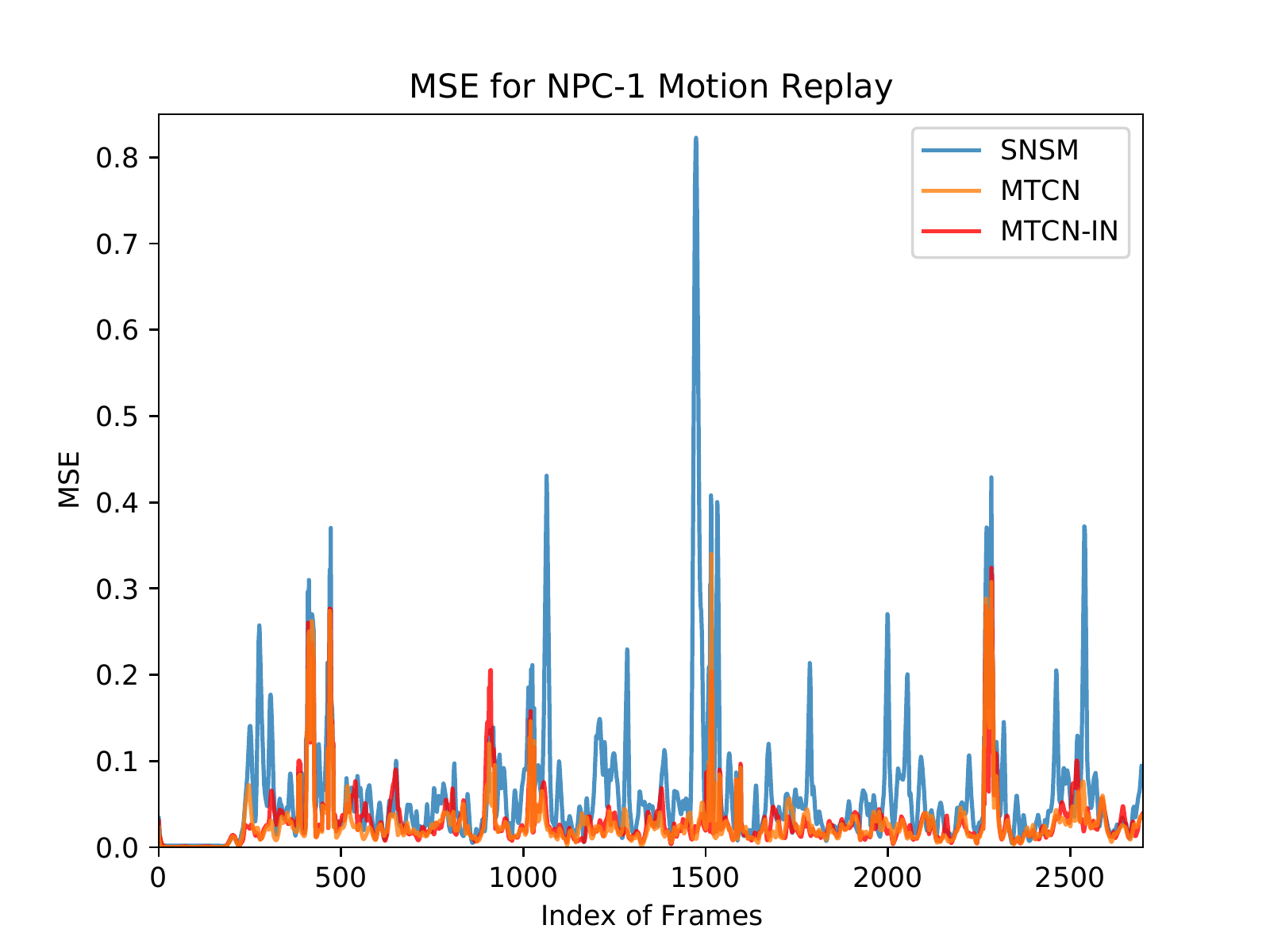}
\caption{\small{The error timeseries for a specific BVH for different sequence models. The horizontal axis is the time stamps (frames), and the vertical axis is the error.}}
\label{fig:mse}
\end{figure}

\begin{table}[t]
    \centering
    \scalebox{0.9}{
    \begin{tabular}{|c||c|c|c|}
    \hline
         Style  & SNSM  & MTCN & MTCN-IN \\
     \hline
    NPC-1   & 0.0772  & \textbf{0.0268}  & 0.0303 \\
    NPC-2   & 0.2080  & 0.1068  & \textbf{0.1022} \\
    NPC-3   & 0.4443  & \textbf{0.1120}  & 0.2309 \\
    NPC-Gun & 0.0300  & 0.0138  & \textbf{0.0137} \\
    FEM-1   & 0.1876  & \textbf{0.0796} & 0.0977 \\
    FEM-2   & 0.1989  & \textbf{0.1332} & 0.1366 \\
    FEM-3   & 0.2529  & \textbf{0.1202} & 0.1957 \\
    FEM-4   & 0.0887  & \textbf{0.0640} & 0.0656 \\
    \hline
    \end{tabular}
    }
    \caption{\small{MSE for Sequence Models on walking motions.}}
    \label{tab:mse}
\end{table}

\begin{figure*}[t!]
\centering
\includegraphics[width=1\linewidth]{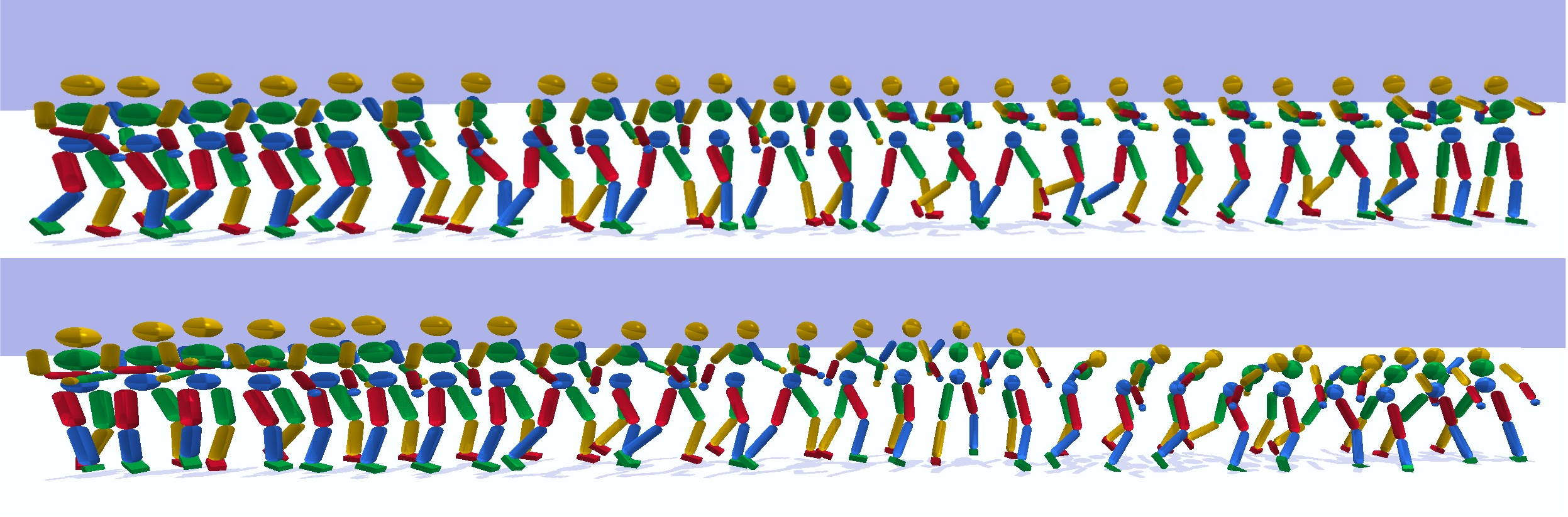}
\caption{\small{The series of motions represent the online style transferring process of the model MTCN. From left to right, the upper image shows style changing from FEM-3 to FEM-4, then to NPC-GUN and stand at last. The lower figure shows motions styles varying from Fem-1 to Fem-2, then to NPC-2 and stand still.}}
\label{fig:mtcn_online}
\end{figure*}

As is suggested in the Fig.~\ref{fig:mse} and Table~\ref{tab:mse}, MTCN significantly outperforms the baseline model SNSM. When looking into the replayed motions, we found two reasons for this improvement of performance. First, the SNSM sometimes produces motions that have slight lags in terms of the phases. This would further lead to lags in the future frames, thus causing a misalignment of phases between the generated motions and the ground truth. In Fig.~\ref{fig:mse}, the fluctuation of errors is mainly caused by this lagging effect. Second, compared to MTCN, SNSM generates motions with lower accuracy. This can also be associated with the phase inputs to the gating network. The reason is that in a complex and diverse motion clip, even if the phase $p_t$ of the motions is the same, the motions could have natural variations. The simple phase inputs are unable to capture these. On the other hand, the learned motion features provide MTCN with better descriptions of the motion at the current frame (as an improved substitution of the phase $p_t$), as well as stronger capacity of capturing the intrinsic variations in the data.

Consequently, at the frame where the misalignment is likely to happen, MTCN possesses better performance. Moreover, the overall error of MTCN is much lower than that of SNSM. This suggests that MTCN has decent accuracy of reproducing learned motions, which would benefit the diversity of the online motion generation. On the other hand, SNSM performs poorly in the multi-style setting, which indicates the quality of the generated motions is not adequate for gaming and any other style-related operations.  

As is discussed in Sec.~\ref{sec:mtcn}, MTCN-IN uses WIN for better elimination of style information in the past motion sequence. This certainly causes performance degradation. But as suggested in Table~\ref{tab:mse}, comparatively this degradation is small in most of the styles. 

\subsection{Online Style Transfer} \label{sec:transfer exp}

\begin{table}[t]
    \centering
    \begin{tabular}{|c|c|c|c|}
    \hline
         Ability  & SNSM & MTCN & MTCN-IN  \\
     \hline
    Replay BVH & Poor & Good    & Good    \\
    Online Style Transition & Yes  & Yes    & Yes    \\
    Style Interpolation & Yes  & No    & Yes    \\
    \hline
    \end{tabular}
    \caption{Different Abilities of Models.}
    \label{tab:qual}
\end{table}

In real game design, it is often required to combine different styles or transiting between them. In Sec.~\ref{sec:replay exp}, we only evaluate the accuracy of reproducing the training data, but leave the transitions between styles untested. This requires the model to produce a series of intermediate motions between two styles, and continuously change between them. The ability to interpolate between styles is further evaluated in Sec.~\ref{sec:interpolate exp}. In this section, we focus on the continuous transition itself.

Here we device a qualitative evaluation for the models. The evaluation tests whether the model can flexibly transit between two arbitrary styles that it was trained on. Suppose the user assigns the transition from style $s_1$ to $s_2$. Once the control is issued, we linearly interpolate to change from $s_1$ to $s_2$. More specifically, we have a factor $\lambda$, which mixes the embeddings $s_1$ and $s_2$ to be $s_M = (1-\lambda) s_1 + \lambda s_2$. The $\lambda$ gradually increase from $0$ to $1$ as time goes on. The generated $s_M$ is set to be the style embedding for the model at that frame. We set the whole process to happen within $1$ second. Then, we observe whether the transition happens successfully within this $1$ second.

All the models can successfully perform the arbitrary (shown in Tab.~\ref{tab:qual}). In Fig.~\ref{fig:mtcn_online}, we have two images showing the online style transfer process using MTCN model. From left to right, the the motion sequence frames varying from FEM-3 to FEM-4, then to NPC-GUN and stand still for upper image. Lower figure shows style transfer from FEM-1 to FEM-2 then to NPC-2 and stand at last. 

\subsection{Style Interpolation} \label{sec:interpolate exp}

Finally, we further evaluate the ability of the models to interpolate between different styles, to create a new style. This is similar to the evaluation in \cite{aberman2020unpaired}, but they did it in an offline fashion.

This evaluation is also qualitative. We choose two styles $s_1$ and $s_2$ from the training data, and linearly interpolated between their one-hot embeddings (each 50\%), which produce $s_I = {(s_1 + s_2)}/{2}$. Then, we directly use the interpolated $s_I$ at each frame for the autoregressive motion generation. We then specify the user controls and observe the motions from the new style $s_I$. 

For this task, only the SNSM and MTCN-IN succeed (shown in Tab.~\ref{tab:qual}). This is expected for the reasons mentioned in Sec.~\ref{sec:mtcn}. Unlike MTCN-IN, which depends on the content sequence, MTCN has strong dependency on the raw motions. This makes it hard for MTCN to generate the motions based on a new sequence of motions, which is from the the interpolated style.

Overall, MTCN-IN succeeds as the model with both accurate motion replay and the ability to flexibly transit between and interpolate styles.

\section{Conclusion} \label{sec:conclusion}
In this work, we target the online motion style transfer task with interactive control. We proposed a novel neural network model, namely MTCN. It could handle large number of motion types and extract motion features using TCN feature extractors, which replace the commonly used phase features.

We also propose a baseline model, namely SNSM, which extends the NSM model to accommodate different styles. To make MTCN better transit between and interpolate styles, we augment it to be MTCN-IN. Both MTCN and MTCN-IN have significant improvement over SNSM in terms of motion replay error. MTCN-IN turns out to have the most satisfying performance in our online motion style transfer evaluation pipeline. It is more flexible than MTCN, with small loss of accuracy. MTCN-IN could be directly used for multi-style motion generation in real time applications like games.

%

\clearpage
\bibliographystyle{ACM-Reference-Format}
\bibliography{sample-base}


\begin{thebibliography}{21}


\ifx \showCODEN    \undefined \def \showCODEN     #1{\unskip}     \fi
\ifx \showDOI      \undefined \def \showDOI       #1{#1}\fi
\ifx \showISBNx    \undefined \def \showISBNx     #1{\unskip}     \fi
\ifx \showISBNxiii \undefined \def \showISBNxiii  #1{\unskip}     \fi
\ifx \showISSN     \undefined \def \showISSN      #1{\unskip}     \fi
\ifx \showLCCN     \undefined \def \showLCCN      #1{\unskip}     \fi
\ifx \shownote     \undefined \def \shownote      #1{#1}          \fi
\ifx \showarticletitle \undefined \def \showarticletitle #1{#1}   \fi
\ifx \showURL      \undefined \def \showURL       {\relax}        \fi
\providecommand\bibfield[2]{#2}
\providecommand\bibinfo[2]{#2}
\providecommand\natexlab[1]{#1}
\providecommand\showeprint[2][]{arXiv:#2}

\bibitem[\protect\citeauthoryear{Aberman, Weng, Lischinski, Cohen-Or, and
  Chen}{Aberman et~al\mbox{.}}{2020}]%
        {aberman2020unpaired}
\bibfield{author}{\bibinfo{person}{Kfir Aberman}, \bibinfo{person}{Yijia Weng},
  \bibinfo{person}{Dani Lischinski}, \bibinfo{person}{Daniel Cohen-Or}, {and}
  \bibinfo{person}{Baoquan Chen}.} \bibinfo{year}{2020}\natexlab{}.
\newblock \showarticletitle{Unpaired motion style transfer from video to
  animation}.
\newblock \bibinfo{journal}{\emph{ACM Transactions on Graphics (TOG)}}
  \bibinfo{volume}{39}, \bibinfo{number}{4} (\bibinfo{year}{2020}),
  \bibinfo{pages}{64--1}.
\newblock


\bibitem[\protect\citeauthoryear{Bengio, Vinyals, Jaitly, and Shazeer}{Bengio
  et~al\mbox{.}}{2015}]%
        {bengio2015scheduled}
\bibfield{author}{\bibinfo{person}{Samy Bengio}, \bibinfo{person}{Oriol
  Vinyals}, \bibinfo{person}{Navdeep Jaitly}, {and} \bibinfo{person}{Noam
  Shazeer}.} \bibinfo{year}{2015}\natexlab{}.
\newblock \showarticletitle{Scheduled sampling for sequence prediction with
  recurrent neural networks}.
\newblock \bibinfo{journal}{\emph{arXiv preprint arXiv:1506.03099}}
  (\bibinfo{year}{2015}).
\newblock


\bibitem[\protect\citeauthoryear{B{\"u}ttner and Clavet}{B{\"u}ttner and
  Clavet}{2015}]%
        {buttner2015motion}
\bibfield{author}{\bibinfo{person}{Michael B{\"u}ttner} {and}
  \bibinfo{person}{Simon Clavet}.} \bibinfo{year}{2015}\natexlab{}.
\newblock \showarticletitle{Motion Matching-The Road to Next Gen Animation}.
\newblock \bibinfo{journal}{\emph{Proc. of Nucl. ai}} (\bibinfo{year}{2015}).
\newblock


\bibitem[\protect\citeauthoryear{Gatys, Ecker, and Bethge}{Gatys
  et~al\mbox{.}}{2016}]%
        {gatys2016image}
\bibfield{author}{\bibinfo{person}{Leon~A Gatys}, \bibinfo{person}{Alexander~S
  Ecker}, {and} \bibinfo{person}{Matthias Bethge}.}
  \bibinfo{year}{2016}\natexlab{}.
\newblock \showarticletitle{Image style transfer using convolutional neural
  networks}. In \bibinfo{booktitle}{\emph{Proceedings of the IEEE conference on
  computer vision and pattern recognition}}. \bibinfo{pages}{2414--2423}.
\newblock


\bibitem[\protect\citeauthoryear{Goodfellow, Pouget-Abadie, Mirza, Xu,
  Warde-Farley, Ozair, Courville, and Bengio}{Goodfellow et~al\mbox{.}}{2014}]%
        {goodfellow2014generative}
\bibfield{author}{\bibinfo{person}{Ian~J Goodfellow}, \bibinfo{person}{Jean
  Pouget-Abadie}, \bibinfo{person}{Mehdi Mirza}, \bibinfo{person}{Bing Xu},
  \bibinfo{person}{David Warde-Farley}, \bibinfo{person}{Sherjil Ozair},
  \bibinfo{person}{Aaron Courville}, {and} \bibinfo{person}{Yoshua Bengio}.}
  \bibinfo{year}{2014}\natexlab{}.
\newblock \showarticletitle{Generative adversarial networks}.
\newblock \bibinfo{journal}{\emph{arXiv preprint arXiv:1406.2661}}
  (\bibinfo{year}{2014}).
\newblock


\bibitem[\protect\citeauthoryear{Holden, Habibie, Kusajima, and Komura}{Holden
  et~al\mbox{.}}{2017a}]%
        {holden2017fast}
\bibfield{author}{\bibinfo{person}{Daniel Holden}, \bibinfo{person}{Ikhsanul
  Habibie}, \bibinfo{person}{Ikuo Kusajima}, {and} \bibinfo{person}{Taku
  Komura}.} \bibinfo{year}{2017}\natexlab{a}.
\newblock \showarticletitle{Fast neural style transfer for motion data}.
\newblock \bibinfo{journal}{\emph{IEEE computer graphics and applications}}
  \bibinfo{volume}{37}, \bibinfo{number}{4} (\bibinfo{year}{2017}),
  \bibinfo{pages}{42--49}.
\newblock


\bibitem[\protect\citeauthoryear{Holden, Komura, and Saito}{Holden
  et~al\mbox{.}}{2017b}]%
        {holden2017phase}
\bibfield{author}{\bibinfo{person}{Daniel Holden}, \bibinfo{person}{Taku
  Komura}, {and} \bibinfo{person}{Jun Saito}.}
  \bibinfo{year}{2017}\natexlab{b}.
\newblock \showarticletitle{Phase-functioned neural networks for character
  control}.
\newblock \bibinfo{journal}{\emph{ACM Transactions on Graphics (TOG)}}
  \bibinfo{volume}{36}, \bibinfo{number}{4} (\bibinfo{year}{2017}),
  \bibinfo{pages}{1--13}.
\newblock


\bibitem[\protect\citeauthoryear{Huang and Belongie}{Huang and
  Belongie}{2017}]%
        {huang2017arbitrary}
\bibfield{author}{\bibinfo{person}{Xun Huang} {and} \bibinfo{person}{Serge
  Belongie}.} \bibinfo{year}{2017}\natexlab{}.
\newblock \showarticletitle{Arbitrary style transfer in real-time with adaptive
  instance normalization}. In \bibinfo{booktitle}{\emph{Proceedings of the IEEE
  International Conference on Computer Vision}}. \bibinfo{pages}{1501--1510}.
\newblock


\bibitem[\protect\citeauthoryear{Johnson, Alahi, and Fei-Fei}{Johnson
  et~al\mbox{.}}{2016}]%
        {johnson2016perceptual}
\bibfield{author}{\bibinfo{person}{Justin Johnson}, \bibinfo{person}{Alexandre
  Alahi}, {and} \bibinfo{person}{Li Fei-Fei}.} \bibinfo{year}{2016}\natexlab{}.
\newblock \showarticletitle{Perceptual losses for real-time style transfer and
  super-resolution}. In \bibinfo{booktitle}{\emph{European conference on
  computer vision}}. Springer, \bibinfo{pages}{694--711}.
\newblock


\bibitem[\protect\citeauthoryear{Kingma and Ba}{Kingma and Ba}{2014}]%
        {kingma2014adam}
\bibfield{author}{\bibinfo{person}{Diederik~P Kingma} {and}
  \bibinfo{person}{Jimmy Ba}.} \bibinfo{year}{2014}\natexlab{}.
\newblock \showarticletitle{Adam: A method for stochastic optimization}.
\newblock \bibinfo{journal}{\emph{arXiv preprint arXiv:1412.6980}}
  (\bibinfo{year}{2014}).
\newblock


\bibitem[\protect\citeauthoryear{Kovar, Gleicher, and Pighin}{Kovar
  et~al\mbox{.}}{2008}]%
        {kovar2008motion}
\bibfield{author}{\bibinfo{person}{Lucas Kovar}, \bibinfo{person}{Michael
  Gleicher}, {and} \bibinfo{person}{Fr{\'e}d{\'e}ric Pighin}.}
  \bibinfo{year}{2008}\natexlab{}.
\newblock \showarticletitle{Motion graphs}.
\newblock In \bibinfo{booktitle}{\emph{ACM SIGGRAPH 2008 classes}}.
  \bibinfo{pages}{1--10}.
\newblock


\bibitem[\protect\citeauthoryear{Mason, Starke, and Komura}{Mason
  et~al\mbox{.}}{2022}]%
        {mason2022real}
\bibfield{author}{\bibinfo{person}{Ian Mason}, \bibinfo{person}{Sebastian
  Starke}, {and} \bibinfo{person}{Taku Komura}.}
  \bibinfo{year}{2022}\natexlab{}.
\newblock \showarticletitle{Real-Time Style Modelling of Human Locomotion via
  Feature-Wise Transformations and Local Motion Phases}.
\newblock \bibinfo{journal}{\emph{arXiv preprint arXiv:2201.04439}}
  (\bibinfo{year}{2022}).
\newblock


\bibitem[\protect\citeauthoryear{Oord, Dieleman, Zen, Simonyan, Vinyals,
  Graves, Kalchbrenner, Senior, and Kavukcuoglu}{Oord et~al\mbox{.}}{2016}]%
        {oord2016wavenet}
\bibfield{author}{\bibinfo{person}{Aaron van~den Oord}, \bibinfo{person}{Sander
  Dieleman}, \bibinfo{person}{Heiga Zen}, \bibinfo{person}{Karen Simonyan},
  \bibinfo{person}{Oriol Vinyals}, \bibinfo{person}{Alex Graves},
  \bibinfo{person}{Nal Kalchbrenner}, \bibinfo{person}{Andrew Senior}, {and}
  \bibinfo{person}{Koray Kavukcuoglu}.} \bibinfo{year}{2016}\natexlab{}.
\newblock \showarticletitle{Wavenet: A generative model for raw audio}.
\newblock \bibinfo{journal}{\emph{arXiv preprint arXiv:1609.03499}}
  (\bibinfo{year}{2016}).
\newblock


\bibitem[\protect\citeauthoryear{Peng, Abbeel, Levine, and van~de Panne}{Peng
  et~al\mbox{.}}{2018}]%
        {peng2018character}
\bibfield{author}{\bibinfo{person}{Xue~Bin Peng}, \bibinfo{person}{Pieter
  Abbeel}, \bibinfo{person}{Sergey Levine}, {and} \bibinfo{person}{Michiel
  van~de Panne}.} \bibinfo{year}{2018}\natexlab{}.
\newblock \showarticletitle{DeepMimic: Example-Guided Deep Reinforcement
  Learning of Physics-Based Character Skills}.
\newblock \bibinfo{journal}{\emph{ACM Trans. Graph.}} \bibinfo{volume}{37},
  \bibinfo{number}{4}, Article \bibinfo{articleno}{143} (\bibinfo{date}{jul}
  \bibinfo{year}{2018}), \bibinfo{numpages}{14}~pages.
\newblock
\showISSN{0730-0301}
\urldef\tempurl%
\url{https://doi.org/10.1145/3197517.3201311}
\showDOI{\tempurl}


\bibitem[\protect\citeauthoryear{Smith, Cao, Neff, and Wang}{Smith
  et~al\mbox{.}}{2019}]%
        {smith2019efficient}
\bibfield{author}{\bibinfo{person}{Harrison~Jesse Smith}, \bibinfo{person}{Chen
  Cao}, \bibinfo{person}{Michael Neff}, {and} \bibinfo{person}{Yingying Wang}.}
  \bibinfo{year}{2019}\natexlab{}.
\newblock \showarticletitle{Efficient neural networks for real-time motion
  style transfer}.
\newblock \bibinfo{journal}{\emph{Proceedings of the ACM on Computer Graphics
  and Interactive Techniques}} \bibinfo{volume}{2}, \bibinfo{number}{2}
  (\bibinfo{year}{2019}), \bibinfo{pages}{1--17}.
\newblock


\bibitem[\protect\citeauthoryear{Srivastava, Hinton, Krizhevsky, Sutskever, and
  Salakhutdinov}{Srivastava et~al\mbox{.}}{2014}]%
        {JMLR:v15:srivastava14a}
\bibfield{author}{\bibinfo{person}{Nitish Srivastava},
  \bibinfo{person}{Geoffrey Hinton}, \bibinfo{person}{Alex Krizhevsky},
  \bibinfo{person}{Ilya Sutskever}, {and} \bibinfo{person}{Ruslan
  Salakhutdinov}.} \bibinfo{year}{2014}\natexlab{}.
\newblock \showarticletitle{Dropout: A Simple Way to Prevent Neural Networks
  from Overfitting}.
\newblock \bibinfo{journal}{\emph{Journal of Machine Learning Research}}
  \bibinfo{volume}{15}, \bibinfo{number}{56} (\bibinfo{year}{2014}),
  \bibinfo{pages}{1929--1958}.
\newblock
\urldef\tempurl%
\url{http://jmlr.org/papers/v15/srivastava14a.html}
\showURL{%
\tempurl}


\bibitem[\protect\citeauthoryear{Starke, Zhang, Komura, and Saito}{Starke
  et~al\mbox{.}}{2019}]%
        {starke2019neural}
\bibfield{author}{\bibinfo{person}{Sebastian Starke}, \bibinfo{person}{He
  Zhang}, \bibinfo{person}{Taku Komura}, {and} \bibinfo{person}{Jun Saito}.}
  \bibinfo{year}{2019}\natexlab{}.
\newblock \showarticletitle{Neural state machine for character-scene
  interactions.}
\newblock \bibinfo{journal}{\emph{ACM Trans. Graph.}} \bibinfo{volume}{38},
  \bibinfo{number}{6} (\bibinfo{year}{2019}), \bibinfo{pages}{209--1}.
\newblock


\bibitem[\protect\citeauthoryear{Starke, Zhao, Komura, and Zaman}{Starke
  et~al\mbox{.}}{2020}]%
        {starke2020local}
\bibfield{author}{\bibinfo{person}{Sebastian Starke}, \bibinfo{person}{Yiwei
  Zhao}, \bibinfo{person}{Taku Komura}, {and} \bibinfo{person}{Kazi Zaman}.}
  \bibinfo{year}{2020}\natexlab{}.
\newblock \showarticletitle{Local motion phases for learning multi-contact
  character movements}.
\newblock \bibinfo{journal}{\emph{ACM Transactions on Graphics (TOG)}}
  \bibinfo{volume}{39}, \bibinfo{number}{4} (\bibinfo{year}{2020}),
  \bibinfo{pages}{54--1}.
\newblock


\bibitem[\protect\citeauthoryear{Ulyanov, Vedaldi, and Lempitsky}{Ulyanov
  et~al\mbox{.}}{2016}]%
        {ulyanov2016instance}
\bibfield{author}{\bibinfo{person}{Dmitry Ulyanov}, \bibinfo{person}{Andrea
  Vedaldi}, {and} \bibinfo{person}{Victor Lempitsky}.}
  \bibinfo{year}{2016}\natexlab{}.
\newblock \showarticletitle{Instance normalization: The missing ingredient for
  fast stylization}.
\newblock \bibinfo{journal}{\emph{arXiv preprint arXiv:1607.08022}}
  (\bibinfo{year}{2016}).
\newblock


\bibitem[\protect\citeauthoryear{Yumer and Mitra}{Yumer and Mitra}{2016}]%
        {yumer2016spectral}
\bibfield{author}{\bibinfo{person}{M~Ersin Yumer} {and}
  \bibinfo{person}{Niloy~J Mitra}.} \bibinfo{year}{2016}\natexlab{}.
\newblock \showarticletitle{Spectral style transfer for human motion between
  independent actions}.
\newblock \bibinfo{journal}{\emph{ACM Transactions on Graphics (TOG)}}
  \bibinfo{volume}{35}, \bibinfo{number}{4} (\bibinfo{year}{2016}),
  \bibinfo{pages}{1--8}.
\newblock


\bibitem[\protect\citeauthoryear{Zhang, Starke, Komura, and Saito}{Zhang
  et~al\mbox{.}}{2018}]%
        {zhang2018mode}
\bibfield{author}{\bibinfo{person}{He Zhang}, \bibinfo{person}{Sebastian
  Starke}, \bibinfo{person}{Taku Komura}, {and} \bibinfo{person}{Jun Saito}.}
  \bibinfo{year}{2018}\natexlab{}.
\newblock \showarticletitle{Mode-adaptive neural networks for quadruped motion
  control}.
\newblock \bibinfo{journal}{\emph{ACM Transactions on Graphics (TOG)}}
  \bibinfo{volume}{37}, \bibinfo{number}{4} (\bibinfo{year}{2018}),
  \bibinfo{pages}{1--11}.
\newblock


\end{thebibliography}

%
\end{document}